\documentstyle[aps]{revtex}
\begin{document}
\unitlength=1.0cm
\title{Long-Range Coulomb Interaction and Frequency Dependence
of Shot Noise in Mesoscopic Diffusive Contacts} 
\author{K.\ E.\ Nagaev}
\maketitle

\bigskip
\noindent
{\it 
Institute of Radioengineering and Electronics, Russian Academy
of Sciences, Mokhovaya ul.\ 11, 103907 Moscow, Russia}

\bigskip
\begin{abstract}
The frequency dependence of shot noise in mesoscopic diffusive
contacts is calculated with account taken of long-range Coulomb
interaction and external screening. While the 
low-frequency noise is $1/3$ of noise of
classical Poisson process independently of the contact shape,
the high-frequency noise tends to the full classical value for
long and narrow contacts because of strong screening by the
surrounding medium. In this case, the
current fluctuations at opposite ends of the contact are
completely independent.
\end{abstract}

\section{INTRODUCTION}

Recently, the shot noise in mesoscopic contacts became a subject
of extensive study.  One of the principal results was that in
contacts with a strong elastic scattering, the low-frequency
shot noise is 1/3 of the full noise of classical Poisson
process.  This result was obtained almost simultaneously by
different authors using different methods and, more importantly,
different physical assumptions.  Beenakker and B\"{u}ttiker
\cite{But92}
obtained this result using the multichannel scattering-matrix 
formalism and the
assumption of quantum-coherent transport.  In contrast to this, in
paper \cite{Nag92} this result was obtained using quasiclassical
kinetic equation under the assumption of completely incoherent
local scattering. Although all the impurity-scattering events were
considered to be independent, at low frequencies the 
current-conservation law
resulted in averaging of the random current over the
contact volume, thus establishing a sort of correlation in
electron transport.  In a sense, this appears to be equivalent
to quantum-coherent scattering, in which case the whole impurity
system
is considered as a single scatterer.  In both approaches, the
low-frequency shot noise appears to be independent of contact
geometry \cite{But92}, \cite{Nag92}, \cite{YuNaz}.  However, this
should not be the case for the finite-frequency noise because
it should be affected by long-range Coulomb interactions. In
particular, it should depend on the possibility for the charge
to pile up in the contact, i.e., on its external capacity.
The problem of frequency-dependent noise in diffusive mesoscopic
contacts was addressed by Altshuler, Levitov, and Yakovets
\cite{Altsh} for the case of
quantum-coherent transport, where the frequency dependence of
noise was determined by fermionic
correlations, but no account of long-range Coulomb
interactions was taken in this paper.

In the present paper, we consider the effects of contact geometry
on the frequency dependence of shot noise within the
semiclassical incoherent-scattering approach. We consider the
case where all its dimensions are much larger than the screening
length $\lambda_0$.  The contact of length $L$ is either a
cylinder of circular section with a diameter $2r_0$ or a 
plane-parallel layer of thickness $d_0$ consisting of a metal
with a high impurity
content (see Fig.\  \ref{contact}).  The electrodes are of the
same section, yet the resistivity of their material is
negligible.  The contact is embedded in a perfectly conducting
grounded medium, which is separated from its surface by a thin
insulating film of thickness $\delta_0$ and the dielectric
constant $\varepsilon_d$.  As will be shown below,
this particular choise of contact geometry allows us to avoid
solving Poisson equation in the surrounding medium and reduces
the effects of environmental screening to frequency-dependent 
boundary conditions.

The external circuit is assumed to have a large grounding
capacity, which allows accumulation of the charge in it.

Our consideration is based on the Boltzmann - Langevin approach
first proposed in \cite{Kog}.  The nonuniform extraneous currents
caused by randomness of electron-impurity scattering result in
local charge-density fluctuations in the bulk of the contact.
These fluctuations are effectively screened by the surface charge
induced at the outer surface of the contact and in the
surrounding medium (environmental screening) and at the contact -
electrode interfaces (electrode screening).  As will be shown
below, the finite-frequency noise essentially depends on the
dominating type of screening.

\section{BASIC EQUATIONS}

In the Boltzmann-Langevin approach, the long-range Coulomb
interaction is taken into account by fluctuations of charge
density $\delta\rho$ and self-consistent fluctuations of
electrical field $\delta{\bf E}$.  Consider the case of strong
and purely elastic scattering.  The Boltzmann-Langevin equation
for fluctuations reads 
  \begin{eqnarray} 
  \left[
     \frac{\partial}{\partial t} 
     + 
     {\bf v}\frac{\partial}{\partial{\bf r}} 
     + 
     e{\bf Ev}\frac{\partial}{\partial \epsilon} 
  \right] \,\delta f 
  + 
  \delta I 
  = 
  -e\,\delta{\bf E\,v} \frac{\partial f}{\partial\epsilon} 
  + 
  \delta J^{ext},\label{BE1} 
  \end{eqnarray}
where $\delta J^{ext}$ is the random extraneous flux.  The
correlation function of these fluxes is given by the expression 
\cite{Kog}
 \begin{eqnarray}
 \left<
      \delta J^{ext}({\bf p}, {\bf r}, t)
      \delta J^{ext}({\bf p'},{\bf r'}, t')
 \right> 
 & = &
 \delta ({\bf r - r'})\delta (t - t')  \nonumber\\
 \Bigl(
     \delta_{pp'}\sum_q 
     \lbrace
          W({\bf p}, {\bf p + q})f({\bf p + q})[1 - f({\bf p})]
          & + &
          W({\bf p + q}, {\bf p})f({\bf p})[1 - f({\bf p + q})] 
     \rbrace\nonumber\\
     -
     \lbrace 
          W({\bf p'}, {\bf p})f({\bf p})[1 - f({\bf p'})]
          & + &  
          W({\bf p}, {\bf p'})f({\ p'})[1 - f({\bf p})] 
     \rbrace
 \Bigr) ,
 \label{BE9}
 \end{eqnarray}
where $W({\bf p}, {\bf p'})$ is the probability of scattering
from state ${\bf p'}$ to state ${\bf p}$.  The fluctuation of
electrical field $\delta{\bf E}$ in the right hand side of
Eqn.\ (\ref{BE1}) is determined from the Maxwell equation
  \begin{equation}
  \nabla\,\delta{\bf E} =
  4\pi\,\delta\rho.
  \label{Maxw}
  \end{equation}
The fluctuations of the charge and current density are given by
the expressions 
  \begin{equation}
  \delta\rho({\bf r}, t) 
  =
  e\int d^3 p\, \delta f ({\bf p}, {\bf r}, t),
  \quad
  \delta{\bf j}({\bf r}, t) 
  =
  e\int d^3 p\, {\bf v} \delta f({\bf p}, {\bf r}, t).
  \label{BE4}
  \end{equation}

Now we proceed to the hydrodynamic approach and obtain a closed 
set of equations for macroscopic quantities $\delta {\bf j}$ and 
$\delta\rho$. As the impurity scattering is strong, we can split the 
fluctuation of distribution function into the parts symmetric and 
antisymmetric in the momentum space.
Then we separate the antisymmetric part of Eqn.\ (\ref{BE1}) from
its symmetric part. Note that the extraneous flux contains only
the antisymmetric part because the electron-impurity scattering
does not change the total number of electrons at a given point
with a given energy.  Multiply the antisymmetric part of (\ref{BE1})
by $e{\bf v}$, integrate it with respect to $d^3 p$,
and then multiply both its parts by the elastic scattering time,
$\tau$.  If the characteristic
times considered are much larger than $\tau$,  one obtains
  \begin{eqnarray}
  \delta{\bf j} 
  =
  -D\frac{\partial}{\partial{\bf r}}\,\delta\rho 
  +
  \sigma\,\delta{\bf E} 
  +
  \delta{\bf j^{ext}},
  \label{B_j}
  \end{eqnarray}
where $D = v^2\tau/3$ is the diffusion coefficient, $\sigma =
e^2 N_F D$ is the conductivity of metal, and 
 \begin{math}
  \delta{\bf j^{ext}} 
  = 
  e\tau\int d^3 p\,{\bf v}\,\delta J^{ext}.
  \end{math}
Integrating the symmetric part of Eqn.\ (\ref{BE1}) with respect 
to momentum, one obtains just the current-conservation law 
 \begin{equation}
  \frac{\partial}{\partial t} \,\delta\rho 
  +
  \nabla\,\delta{\bf j}
  = 
  0.
  \label{BE6}
  \end{equation}

Applying the operator $\nabla$ to both parts of Eqn.\ (\ref{B_j})
and making use of Eqns.\ (\ref{BE6}) and (\ref{Maxw}), one obtains a
closed equation for fluctuations of charge density in the form 
  \begin{equation}
  \left(
       \frac{\partial}{\partial t} 
       -
       D\nabla^2 
       +
       4\pi\sigma
  \right)
  \,\delta\rho
  =
  -\nabla\,\delta{\bf j^{ext}}.
  \label{screen}
  \end{equation}
In the left-hand side of this equation, the second term
describes diffusion of electrons, and the third term describes
the Coulomb screening of fluctuations. In principle,
Eqn.\ (\ref{screen}) may be solved for each particular distribution of
$\delta{\bf j^{ext}}$, and then $\delta{\bf E}$ and $\delta{\bf j}$
may be determined from Eqns.\ (\ref{Maxw}) and (\ref{B_j}),
respectively.
In the static limit, Eqn.\ (\ref{screen}) describes the screening of
an extraneous charge with the standard length $\lambda_0$ given by
 $$
 \lambda_0^{-2} =
 \frac{4\pi\sigma}{D} =
 4\pi e^2 N_F.
 $$
To complete the derivation, we must obtain the correlation
function of extraneous currents $\delta{\bf j^{ext}}$. 
As we are restricted to the case of strong impurity scattering,
the distribution function may be considered as isotropic in the
momentum space and dependent only on the coordinate {\bf r} and
energy $\epsilon$. Multiply Eqn.\ (\ref{BE9}) by $e\tau v_{\alpha}$
and $e\tau v'_{\beta}$, where $\alpha$ and $\beta$ label vector
components, and integrate it with respect to $d^3 p$ and $d^3
p'$. As a result, one obtains the spectral density of extraneous
currents in the form
 \begin{equation}
 \left<
 \delta j_{\alpha}^{ext}({\bf r})
 \delta j_{\beta}^{ext}({\bf r'})
 \right>_{\omega}
 =
 4\sigma\delta_{\alpha \beta} \delta({\bf r - r'})
 \int d\epsilon\,f(\epsilon, {\bf r})[1 - f(\epsilon, {\bf r})].
 \label{S_j general}
 \end{equation}

Because of smallness of $\lambda_0$, the relationship between
the extraneous currents and fluctuations of charge density in
the bulk of the sample may be considered as local. Taking the
Fourier transform of Eqn.\ (\ref{screen}) with respect to time,
integrating it over the space, and making use of the Gauss
theorem, one obtains:
 \begin{equation}
 \delta\rho
 =
 -(-i\omega + 4\pi\sigma)^{-1}
 \nabla\delta{\bf j^{ext}}
 \label{delta rho}
 \end{equation} 

Note that the quasineutrality condition does not hold for
fluctuations. 
Introduce the fluctuating potential $\delta\phi$ that satisfies 
the Poisson equation 
  \begin{equation}
  \nabla^2\delta\phi
  =
  -4\pi\delta\rho.
  \label{Poisson}
  \end{equation}
Consider the boundary conditions for $\delta\phi$ at the outer
insulated surface of the contact. The normal derivatives of
$\delta\phi$ in the dielectric layer and inside the metal are
related by the expression

  \begin{equation}
  \left.
     \varepsilon_d \frac{\partial\delta\phi}{\partial n}
  \right|_d
  -
  \left.
     \frac{\partial\delta\phi}{\partial n}
  \right|_s
  =
  -4\pi\delta\sigma_s,
  \label{GE3}
  \end{equation}
where $\delta\sigma_s$ is fluctuating surface charge density
induced by the extraneous currents. On the other hand, this
charge density satisfies the charge-balance equation
  \begin{equation}
  -i\omega\delta\sigma_s
  =
  -\sigma\left.
  \frac{\partial\delta\phi}{\partial n}
  \right|_s.
  \label{GE4}
  \end{equation}
As the thickness of dielectric layer is much smaller than the
size of the contact, the electric field across it may be
considered uniform so that 
$\left.\partial\delta\phi/\partial n\right|_d = -\delta_0^{-1}
\left.\delta\phi\right|_s$. With this condition, Eqns.\
(\ref{GE3}) and (\ref{GE4}) give the boundary condition for
$\delta\phi$ in the form
  \begin{equation}
  \left.\left[
  -i\omega\varepsilon_d\delta_0^{-1}\delta\phi
  + (-i\omega + 4\pi\sigma)
  \frac{\partial\delta\phi}{\partial n}
  \right]\right|_s
  = 0.
  \label{outer bound}
  \end{equation}
It is easily seen that at $\omega = 0$, Eqn.\ (\ref{outer bound}) takes
the form $\left.\partial\delta\phi/\partial n\right|_s = 0$,
while at $\omega\to\infty$, it takes the form
$\left.\delta\phi\right|_s = 0$.

As the voltage drop across the contact is held constant,
fluctuations of potential are zero at the contact-electrode
interfaces: 
  \begin{equation}
  \left.\delta\phi\right|_i = 0.
  \label{GE6}
  \end{equation}
As the electrodes are perfect conductors, $\partial\phi/\partial
n = 0$ inside them. Equation (\ref{GE3}) holds for
contact-electrode interfaces, but the charge-balance equation
takes now the form
  \begin{equation}
  -i\omega\delta\sigma_s 
  =
  -\sigma\left.
  \frac{\partial\delta\phi}{\partial n}
  \right|_i
  -
  \delta j_n,
  \label{GE7}
  \end{equation}
where $\delta j_n$ is the fluctuation of current flowing into 
the electrodes
from the contact. From Eqns.\ (\ref{GE3}) and (\ref{GE7}), it
follows that the density of outgoing current is given by
  \begin{equation}
  \delta j_n
  =
  \left(
  \frac{i\omega}{4\pi} - \sigma
  \right)
  \left.
  \frac{\partial\delta\phi}{\partial n}
  \right|_i.
  \label{outgoing current}
  \end{equation}

From the standpoint of average current, the problem is purely
one-dimensional, so the average distribution function
$f(\epsilon, x)$ obeys the one-dimensional diffusion equation,
its boundary values being zero-temperature Fermi distribution
functions shifted in energy by $eV$ with respect to each other.
As the contact is much shorter than the characteristic inelastic
length,
  \begin{equation}
  f(\epsilon,x) =
  \left\{
  \begin{array}{ll}
  0,       & \epsilon > eV/2\\
  1 - x/L, & eV/2 > \epsilon > -eV/2\\
  1,       & \epsilon
 < -eV/2.
  \end{array}
  \right.
  \label{f average}
  \end{equation}
With this distribution function, the expression for the
spectral density of extraneous currents (\ref{S_j general}) 
takes the form
  \begin{equation}
  \left<
  \delta j_{\alpha}^{ext}({\bf r})
  \delta j_{\beta}^{ext}({\bf r'})
  \right>_{\omega}
  =
  4\sigma\delta_{\alpha \beta} \delta({\bf r - r'})
  \frac{x}{L}
  \left(
  1 - \frac{x}{L}
  \right) .
  \label{S_j - 1d}
  \end{equation}
\section{CIRCULAR-SECTION CONTACT. ANALYTICAL RESULTS}

Consider the Poisson equation with the boundary conditions
(\ref{outer bound}). As the system is axially symmetric, all the
quantities may be considered as independent of the azimuthal
angle and dependent only on the longitudinal coordinate $x$ and
radius $r$. In this case, the boundary condition (\ref{outer
bound}) takes the form
  \begin{equation}
  \left.\left(
              r_0\frac{\partial\delta\phi}{\partial r}
              +
              \mu\delta\phi
  \right)\right|_{r = r_0}
  = 0,
  \qquad
  \mu
  \equiv
  \frac{-i\omega}{-i\omega + 4\pi\sigma}
  \frac{\varepsilon_d r_0}{\delta_0}.
  \label{mu}
  \end{equation}

Supposefirst that $\mu$ is real and positive. Then one may 
introduce a system of 
normalized eigenfunctions $\psi_n$ satisfying the equation
  \begin{equation}
  \frac{1}{r}\frac{d}{dr}\left( r\frac{d}{dr}\psi_n(r)\right)
  +
  k_n^2\psi_n(r)
  = 0
  \label{Bessel}
  \end{equation}
with the boundary conditions (\ref{mu}). These functions are 
given by
  \begin{equation}
  \psi_n(r)
  =
  \frac{1}{\pi^{1/2}r_0}
  \frac{J_0(k_nr)}{\sqrt{J_0^2(k_nr_0) + J_1^2(k_nr_0)}},
  \label{eigenfunction}
  \end{equation}
where $J_0$ and $J_1$ are the Bessel functions of zeroth and
first order and the eigenvalues $k_n$ are determined from the 
equation
  \begin{equation}
  k_nr_0\frac{J_1(k_nr_0)}{J_0(k_nr_0)} 
  = 
  \mu.
  \label{eigenvalue}
  \end{equation}
Since functions $\psi_n$ form an orthogonal
basis, for an arbitrary charge-density fluctuation
$\delta\rho$, the solution of Poisson equation (\ref{Poisson})
with the boundary condition (\ref{mu}) is given by the
expression
  \begin{equation}
  \phi(x, r) 
  =
  -4\pi\sum_{n=0}^{\infty}\psi_n(x, r)
  \int\limits_0^L dx'\,g_n(x, x')
  \int dS'\,\psi_n(r')\rho(x', r'),
  \label{delta phi}
  \end{equation}
where $dS' = 2\pi r'dr'$ and $g_n(x, x')$ is the Green's
function of the equation
  \begin{equation}
  \left(
  \frac{d^2}{dx^2} - k_n^2
  \right)
  g_n(x, x')
  =
  \delta(x - x')
  \label{g definition}
  \end{equation}
with the boundary conditions $g_n(0, x') = g_n(L, x') = 0$,
which is given by the expression
  \begin{eqnarray}
  g_n(x<x')
  & = &
  -\frac{\sinh(k_nx)\sinh[k_n(L-x')]}{k_n\sinh(k_nL)},
  \nonumber\\
  g_n(x>x')
  & = &
  -\frac{\sinh(k_nx')\sinh[k_n(L-x)]}{k_n\sinh(k_nL)}.
  \label{g explicit}
  \end{eqnarray}

Equation (\ref{delta phi}) may be analytically continued to
complex values of $\mu$ given by Eqn.\ (\ref{mu}).
Substituting Eqn.\ (\ref{delta rho}) for $\delta\rho$ into Eqn.\
(\ref{delta phi}) for $\delta\phi$ and then (\ref{delta phi})
into (\ref{outgoing current}), one obtains the expression for
the fluctuation of the current flowing through the left end of 
the contact in the form 
  \begin{equation}
  \delta I(0)
  =
  S_0\sum_{n=0}^{\infty}\overline{\psi_n}
  \int\limits_0^L dx'\, \int dS'\,
  \left[
  \left.
    \frac{\partial^2g_n(x, x')}{\partial x \partial x'}
    \right|_{x=0}
  \psi_n(r')\delta j_x^{ext}
  +
  \left.
    \frac{\partial g_n(x, x')}{\partial x}
  \right|_{x=0}
  \frac{\partial\psi_n}{\partial r'}
  \delta j_r^{ext}
  \right],
  \label{delta I}
  \end{equation}
where $S_0 = \pi r_0^2$ and $\overline{\psi_n}$ is $\psi_n$
averaged over the cross section of the contact:
  \begin{equation}
  \overline{\psi_n}
  =
  \frac{1}{S_0}\int dS\,\psi_n(r)
  =
  \frac{2J_1(k_nr_0)}{\pi^{1/2} r_0^2 k_n
    \sqrt{J_0^2(k_nr_0) + J_1^2(k_nr_0}}.
  \label{psi averaged}
  \end{equation}
To obtain the fluctuation of the current flowing through 
the right end of the
contact, $\delta I(L)$, one must substitute $x = L$ for $x =
0$ in Eqn.\ (\ref{delta I}).

At $\omega = 0$, all transverse modes with $n \ne 0$ have vanishing 
cross-sectional averages, $\overline{\psi_n} = 0$, and the 
corresponding longitudinal factors $g_n$ exponentially decay at 
$|x - x'| > r_0$. This is quite natural because the electrical field 
produced by a charge inside the contact cannot penetrte through its
outer surface and is uniformly distributed over the contact cross 
section at large distances from the source. Hence the only contribution 
to Eqn.\ (\ref{delta I}) will be given by the lowest transverse mode with 
$k_0 = 0$ and $\psi_0(r) = \pi^{-1/2}r_0^{-1}$. In this case, Eqn.\ 
(\ref{delta I}) takes the form
  \begin{equation}
  \delta I(0)
  =
  \frac{1}{L}\int\limits_0^L dx \int dS\, \delta j_x^{ext}.
  \label{I - zero frequency}
  \end{equation}
This is just the result obtained in \cite{Nag92}.

Consider now the case where the contact
length $L$ is much larger than its diameter $2r_0$ and the 
frequencies are sufficiently low, i.e., $\omega \ll 
4\pi\sigma\delta_0/\varepsilon_d r_0$. In this case, the corrections
to the zero-frequency eigenfunctions $\psi_n$, as well as the
corrections to the products $k_nr_0$ with $n \ne 0$, are
proportional to $\mu$ and therefore small; hence the
contributions to $\delta I$ (\ref{delta I}) from the modes with
$n \ne 0$ remain insignificant. However, the lowest eigenvalue is 
given by
  \begin{equation}
  k_0 
  =
  r_0^{-1}(2\mu)^{1/2},
  \label{k_0}
  \end{equation}
and the product $k_0L$ may be sufficiently large. Therefore, 
the contribution from the lowest mode governed by $g_0(0, x)$ may 
change significantly. In view of this, the expression for $\delta I$
takes the form
  \begin{equation}
  \delta I(0)
  =
  \int\limits_0^L dx\, k_0
  \frac{\cosh[k_0(L - x)]}{\sinh(k_0L)}
  \int dS\,\delta j_x^{ext},
  \label{delta I - low freq}
  \end{equation}
where $k_0$ is given by Eqn.\ (\ref{k_0}). Physically, this
implies that the contact is represented as an alternating series
of resistors with generators of random current and grounding 
capacities connecting the
electrodes (see Fig.\ \ref{model}). Note that $\delta I(0)$ 
is phase shifted with respect
to the extraneous current inducing it. Multiplying Eqn.\
(\ref{delta I - low freq}) by its complex conjugate,
substituting the spectral density of extraneous currents
(\ref{S_j - 1d}) into the product, and performing the
integration with respect to $x$, one obtains:
  \begin{equation}
  S_I^{LL}(\omega)
  =
  2eI\,
  \left[
       1
       -
       \frac{1}{\gamma_{\omega}L}\,
       \frac{\sinh(2\gamma_{\omega}L) - \sin(2\gamma_{\omega}L)}
            {\cosh(2\gamma_{\omega}L) - \cos(2\gamma_{\omega}L)}
   \right],    
  \label{S_I - low freq}
  \end{equation}
where
  \begin{equation}
  \gamma_{\omega}
  =
  \frac{1}{2}
  \sqrt{\frac{\omega\varepsilon_d}{\pi\sigma\delta_0 r_0}}.
  \label{gamma}
  \end{equation}
The frequency dependence of the shot noise is shown in Fig.\
\ref{long}.
At zero frequency, we rederive the well known result $S_I^{LL} =
\frac{2}{3}eI$. However, at frequencies about $\sigma\delta_0
r_0/\varepsilon_d L^2$, the spectral density sharply rises and
tends to the full value of classical shot noise,
$S_I^{LL} = 2eI$. This suggests that the corresponding
correlation function is negative at $t \ne t'$. The anticorrelation is the
consequence of the Coulomb repulsion of electrons: an entrance
of an electron into the contact decreases for some time the probability
for another electron to enter it, similarly to the case
of a single-electron transistor \cite{Kor et al},
\cite{Korotkov}. 

Along with the spectral density of noise at one
end of the contact, one may also consider the cross-correlated 
spectral density
$$
  S_I^{LR} 
  \equiv 
  \frac{1}{2}
  \left<\delta I(0, \omega) \delta I(L, -\omega)
        +
        \delta I(0, -\omega) \delta I(L, \omega)
  \right>,
$$
which describes the correlation between the currents flowing
through the opposite ends of the contact. Multiplying Eqn.\
(\ref{delta I - low freq}) by its complex conjugate for $\delta
I(L)$ and performing the integration with the spectral density
of extraneous currents (\ref{S_j - 1d}), one obtains:
  \begin{eqnarray}
  S_I^{LR}(\omega)
  =
  \frac{4eI}{\gamma_{\omega}L}\,
  \frac{\cosh(\gamma_{\omega}L)\sin(\gamma_{\omega}L)
        - 
        \cos(\gamma_{\omega}L)\sinh(\gamma_{\omega}L)}
       {\cosh(2\gamma_{\omega}L) - \cos(2\gamma_{\omega}L)}. 
  \label{cross - low freq}
  \end{eqnarray}
The frequency dependence of $S_I^{LR}$ is also shown in Fig.\
\ref{long}. At
$\omega = 0$, it also equals $\frac{2}{3}eI$. However in
contrast to $S_I^{LL}(\omega)$, it sharply decreases with
increasing frequency and tends to zero in an oscillatory way
with further increase of frequency.

Consider now the high-frequency limit. In this case, the
boundary condition (\ref{outer bound}) takes the form
$\psi_n(r_0) = 0$, so that the quantities $k_nr_0$ are the zeros
of zero-order Bessel function. In this case, functions $\psi_n$
are real and form an orthogonal system.
Owing to the orthonormality conditions, the expression for the
spectral density of noise may be written in the form
  \begin{eqnarray}
  S_I^{LL}(\infty)
  =
  4S_0^2eV\sigma\sum_{n=1}^{\infty}\overline{\psi_n}^2 & &
  \int\limits_0^L dx'\,\frac{x'}{L}
  \left(1 - \frac{x'}{L}\right)
\nonumber\\ & &
  \times
  \Biggl\{
     \Biggr[
       \left.
         \frac{\partial^2g_n(x, x')}{\partial x \partial x'} 
       \right|_{x=0}
     \Biggl]^2
     +
     \,k_n^2
     \Biggl[
       \left.
        \frac{\partial g_n(x, x')}{\partial x}
       \right|_{x=0}
     \Biggr]^2
  \Biggl\},
  \label{S_I - inf}  
  \end{eqnarray}

As $J_0(k_nr_0) = 0$, Eqn.\ (\ref{psi averaged}) reduces to
$\overline{\psi_n} = 2\pi^{-1/2}r_0^{-2}k_n^{-1}$. Substituting
the explicit expressions for $g_n$ (\ref{g explicit}) into Eqn.\
(\ref{S_I - inf}), one obtains:
  \begin{equation}
  S_I^{LL}(\infty)
  =
  8eI\sum_{n = 1}^{\infty}\frac{1}{(k_nr_0)^2} \coth(k_nL)
  \left[
     \coth(k_nL) - \frac{1}{k_nL}
  \right],
  \label{LL series}
  \end{equation}
Similarly, one obtains for the cross-correlated spectral
density: 
  \begin{equation}
  S_I^{LR}(\infty)
  =
  8eI\sum_{n = 1}^{\infty}\frac{1}{(k_nr_0)^2}
  \frac{1}{\sinh(k_nL)}
  \left[
     \coth(k_nL) - \frac{1}{k_nL}
  \right].
  \label{LR series}
  \end{equation}
In the limiting case of $r_0 \gg L$, both expressions take the
form 
  \begin{equation}
  S_I^{LL}(\infty) 
  = 
  S_I^{LR}(\infty) 
  = 
  \frac{8}{3}eI \sum_{n = 1}^{\infty}\frac{1}{(k_nr_0)^2}
  =
  \frac{2}{3}eI.
  \label{r >> L}
  \end{equation}
In the opposite limiting case of $r_0 \ll L$, Eqn.\ (\ref{LL
series}) takes the form
  \begin{equation}
  S_I^{LL}(\infty)
  =
  8eI\sum_{n = 1}^{\infty}\frac{1}{(k_nr_0)^2}
  =
  2eI,
  \label{r << L}
  \end{equation}
whereas $S_I^{LR}(\infty)$ (\ref{LR series}) tends to zero
according to the exponential law. The $L/r_0$ dependences of
$S_I^{LL}(\infty)$ and $S_I^{LR}(\infty)$ are shown in Fig.\
\ref{high-freq}.

\section{PLANAR CONTACT. NUMERICAL RESULTS}

Consider now a planar contact in the shape of a layer of
thickness $d_0$ in the $y$ direction ($0 < y < d_0$) and of
width $W$ ($W \gg {\rm max}(d_0, L)$) in the $z$ direction, the 
average current flowing in the $x$ direction. Because of large
$W$, the effects of boundaries in the $z$ direction may be
neglected and all the quatities may be considered as independent
of $z$. Introduce an 
orthonormal system of functions
  \begin{eqnarray}
  \varphi_n(x)
  =& &
  \sqrt{\frac{2}{L}}\sin(q_nx),
  \nonumber\\
  q_n =& & \frac{\pi n}{L},
  \label{sin}
  \end{eqnarray}
which obey the boundary conditions $\varphi_n(0) = \varphi_n(L)
= 0$. For an arbitrary charge-density fluctuation $\delta\rho$,
the potential fluctuation $\delta\phi$ induced by it may be
presented in the form
  \begin{equation}
  \delta\phi(x, y)
  =
  -4\pi\sum_{n = 1}^{\infty}\varphi_n(x)
  \int\limits_0^L dx'\,\varphi(x')
  \int\limits_0^{d_0} dy'\, Q_n(y, y')\,\delta\rho(x', y'),
  \label{phi - rho planar}
  \end{equation}
where $Q_n(y,y')$ satisfies the equation
  \begin{equation}
  \left(
     \frac{\partial^2}{\partial y^2} - q_n^2
  \right)
  Q_n(y, y')
  =
  \delta(y - y')
  \label{Q_n - definition}
  \end{equation}
with the boundary conditions
  \begin{eqnarray}
  \Biggl(
     \frac{-i\omega\varepsilon_d}{4\pi\sigma\delta_0}Q_n
     +
     \frac{\partial Q_n}{\partial y}
  \Biggr)
  \Biggr|_{y = d_0}
  =
  0,\qquad
  \Biggl(
      \frac{i\omega\varepsilon_d}{4\pi\sigma\delta_0}Q_n
      +
      \frac{\partial Q_n}{\partial y}
  \Biggr)
  \Biggr|_{y = 0}\,\,
  =
  0.
  \label{Q_n bound}
  \end{eqnarray}
Explicitly, $Q_n$ for $y > y'$ is given by the expression
  \begin{eqnarray}
  Q_n(y, y')
  = 
  -\frac{1}{2q_n}\,& &
  \frac{q_nd_0\cosh[q_n(d_0 - y)] - i\Omega\sinh[q_n(d_0 - y)]}
       {q_nd_0\cosh(q_nd_0/2) - i\Omega\sinh(q_nd_0/2)}
       \qquad\qquad\qquad\qquad
  \nonumber\\ &  &\qquad\qquad\qquad\qquad\qquad
  \times
  \frac{q_nd_0\cosh(q_ny') - i\Omega\sinh(q_ny')}
       {q_nd_0\sinh(q_nd_0/2) - i\Omega\cosh(q_nd_0/2)},
  \label{Q_n explicit}
  \end{eqnarray}
where $\Omega = \omega\varepsilon_d d_0/4\pi\sigma\delta_0$ is the
dimensionless frequency. The corresponding expression for $y <
y'$ is obtained from Eqn.\ (\ref{Q_n explicit}) by interchanging
$y$ and $y'$. Substituting Eqn.\ (\ref{delta rho}) for
$\delta\rho$ into Eqn.\ (\ref{phi - rho planar}) and then
substituting (\ref{phi - rho planar}) into (\ref{outgoing
current}), one obtains the expression for the fluctuation of 
current flowing through the left end of the contact in the form 
  \begin{eqnarray}
  \delta I(0)
  =
  W\sum_{n = 1}^{\infty}& &
  \left.\frac{d\varphi_n}{dx}\right|_{x=0}
  \int\limits_0^L dx' \int\limits_0^{d_0}dy
  \int\limits_0^{d_0}dy'\,
  \nonumber\\& & 
  \Biggr[\frac{d\varphi_n(x')}{dx'} 
         Q_n(y,y')
         \delta j_x^{ext} 
         +
         \varphi_n(x')
         \frac{\partial Q_n(y, y')}{\partial y'}
         \delta j_y^{ext}
  \Biggl].
  \label{delta I planar}
  \end{eqnarray}
The fluctuation of current flowing through the right end of the
contact may be obtained by substituting $x = L$ for $x = 0$
in Eqn.\ (\ref{delta I planar}). Using the correlator of
extraneous currents (\ref{S_j - 1d}), one obtains the
expressions for the spectral densities $S_I^{LL}$ and $S_I^{LR}$
in the form
  \begin{eqnarray}
  S_I^{LL} 
  =
  8eId_0\sum_{m =1}^{\infty}\sum_{n = 1}^{\infty} & & k_m k_n
  \left(
       M_{mn}P_{mn}'' + M_{mn}'' P_{mn}
  \right), 
  \label{double sum LL}\\
  S_I^{LR}
  =
  4eId_0\sum_{m =1}^{\infty}\sum_{n = 1}^{\infty}
  [(-1)^m + & &(-1)^n] k_m k_n
  \left(
       M_{mn}P_{mn}'' + M_{mn}''P_{mn}
  \right).
  \label{double sum LR}
  \end{eqnarray} 
In these expressions, we used the notation
  \begin{equation}
  M_{mn}
  =
  \frac{1}{d_0^2}\int\limits_0^{d_0}dy
  \int\limits_0^{d_0}dy_1
  \int\limits_0^{d_0}dy_2
  \,
  Q_m(y_1,y)Q_n^*(y_2,y), \label{M_mn def}
  \end{equation}
  \begin{equation}
  M_{mn}''
  =
  \frac{1}{d_0^2}\int\limits_0^{d_0}dy
  \int\limits_0^{d_0}dy_1
  \int\limits_0^{d_0}dy_2
  \,
  \frac{\partial Q_m(y_1, y)}{\partial y}
  \frac{\partial Q_n^*(y_2, y)}{\partial y},
  \label{M'' def}
  \end{equation}
  \begin{equation}
  P_{mn}
  =
  \int\limits_0^L dx\, \varphi_m(x) \varphi_n(x)
  \frac{x}{L}
  \left(
       1 - \frac{x}{L}
  \right),
  \label{P_mn def}
  \end{equation}
  \begin{equation}
  P_{mn}''
  =
  \int\limits_0^L dx\,
  \frac{\partial \varphi_m}{\partial x}
  \frac{\partial \varphi_n}{\partial x}
  \left(
       1 - \frac{x}{L}
  \right),
  \label{P'' def}
  \end{equation}
Using the notation $t_m = \tanh(k_md_0/2)$ and $D_m = 
(k_m^2d_0^2t_m^2 + \Omega^2)^{-1}$ and performing partial
summation over the internal index, one may bring Eqns.\
(\ref{double sum LL}) and (\ref{double sum LR}) to the form
  \begin{eqnarray}
  S_I^{LL}(\omega)
  =
  \frac{2}{3}eI
  +
  4eI\Omega^2\sum_{m = 1}^{\infty}
  \Biggl\{
        \Biggl(
              \frac{2}{3}
              -
              8\frac{1 + (-1)^m}{k_m^2 L^2}
        \Biggr)
        \frac{D_m t_m}{k_m d_0}
        -
        \frac{D_m}{k_m^2 L^2} (1 - t_m^2)
  \nonumber\\
        -
        \frac{8}{L^2}{\sum_{n=1}^{\infty}}'[1 + (-1)^{m+n}]
        \frac{D_m D_n}{d_0(k_m^2 - k_n^2)^2}
        (d_0^2 k_m t_m k_n t_n + \Omega^2)
        (k_m t_m + k_n t_n)
  \Biggr\},
  \label{S^LL expl}
  \end{eqnarray}
  \begin{eqnarray}
  S_I^{LR}(\omega)
  =
  \frac{2}{3}eI
  +
  4eI\Omega^2\sum_{m = 1}^{\infty}(-1)^m
  \Biggl\{
        \Biggl(
              \frac{2}{3}
              -
              8\frac{1 + (-1)^m}{k_m^2 L^2}
        \Biggr)
        \frac{D_m t_m}{k_m d_0}
        -
        \frac{D_m}{k_m^2 L^2} (1 - t_m^2)
  \nonumber\\
        -
        \frac{8}{L^2}{\sum_{n=1}^{\infty}}'[1 + (-1)^{m+n}]
        \frac{D_m D_n}{d_0(k_m^2 - k_n^2)^2}
        (d_0^2 k_m t_m k_n t_n + \Omega^2)
        (k_m t_m + k_n t_n)
  \Biggr\},
  \label{S^LR expl}
  \end{eqnarray}
where the primes by the sums over $n$ show that $n \ne m$.

The contour plots of $S_I^{LL}$ and $S_I^{LR}$ vs.\ logarithms of
frequency and contact length are shown in Figs.\ \ref{contour LL}
and \ref{contour LR}.
Qualitatively, their behavior is similar to that in the case of
a circular-section contact. At low frequencies and small contact
lengths, both quantities tend to $\frac{2}{3}eI$. At high
frequencies and large contact lengths, $S_I^{LL}$ and $S_I^{LR}$
tend to $2eI$ and zero, respectively. It is also clearly seen
that at $L/d_0 \ge 2.86$, the frequency dependences of
$S_I^{LR}$ exhibit negative portions.

\section{CONCLUSION}

Both circular and planar contacts exhibit qualitatively similar
noise properties. At small length-to-width ratios, when the
screening of charge fluctuations by the electrodes is more
efficient than the screening by the ambient medium and pile-up
of the charge in the contact is forbidden, the effects
of long-range Coulomb interaction reduce to averaging the
extraneous currents over the volume of the contact at arbitrary 
frequencies. The
situation is different, however, for long and narrow contacts,
where the charge fluctuations are mostly screened by the ambient
medium and pile-up of charge in the contact is allowed. At
sufficiently high frequencies, the correlation length of
fluctuations becomes smaller than the length of the contact.  In
this case, the fluctuations of current at the ends of the
contact, which are observed in the external circuit, are
dominated by extraneous currents in the narrow adjacent layers.
The corresponding spectral densities are equal to that of the
classical shot noise, $2eI$, while the fluctuations at different
contact ends are completely independent.

This work was supported by DOE's Grant
\#DE-FG02-95ER14575 and by the Russian Foundation for Basic
Research (project \#96-02-16663-a).

The author acknowledges a fruitful discussion with G. B. Lesovik.

\bigskip
\centerline{FIGURE CAPTIONS}

\bigskip
\begin{figure}
\caption{
Longitudinal cross section of the contact. The dotted rectangle
is the metal with impurities, thin solid lines show the contact
- electrode interfaces, thick lines show the dielectric layers
of thickness $\delta_0$,
and the hatched areas show the grounded ambient medium.
}\label{contact}
\end{figure}
\begin{figure}
\caption{
Physical model of noise in a long and narrow contact. Each
section of the $R-C$ line contains a generator of random current.
}\label{model}
\end{figure}
\begin{figure}
\caption{
Dependences of the normalized spectral density of noise at one of
the contact ends $S_I^{LL}/2eI$ (solid line) and the
cross-correlated spectral density $S_I^{LR}/2eI$ (dashed line)
on the dimensionless frequency $\omega L^2\varepsilon_d / 
4\pi\delta_0r_0$ for a long narrow contact.
}\label{long}
\end{figure}
\begin{figure}
\caption{
Dependences of the normalized spectral density of noise at one of
the contact ends $S_I^{LL}/2eI$ (solid line) and cross-correlated 
spectral density $S_I^{LR}/2eI$ (dashed line) on the
length-to-radius ratio $L/r_0$ in the high-frequency limit.
}\label{high-freq}
\end{figure}
\begin{figure}
\caption{
Contour plots of $S_I^{LL}$ vs. logarithms of normalized
frequency $\Omega = \omega\varepsilon_d d_0/4\pi\sigma\delta_0$ and 
normalized length $L/d_0$ for the planar contact.
}\label{contour LL}
\end{figure}
\begin{figure}
\caption{
Contour plots of $S_I^{LR}$ vs. logarithms of normalized
frequency $\Omega = \omega\varepsilon_d d_0/4\pi\sigma\delta_0
$ and 
normalized length $L/d_0$ for the planar contact.
}\label{contour LR}
\end{figure}

\begin{thebibliography}{99}
\bibitem{But92} C. W. J. Beenakker and M. B\"{u}ttiker, Phys. Rev. 
B {\bf 46}, 1889 (1992).
\bibitem{Nag92} K. E. Nagaev, Phys. Lett. A {\bf 169}, 103 (1992).
\bibitem{YuNaz} Yu. Nazarov, Phys. Rev. Lett. {\bf 73}, 134
(1994).
\bibitem{Kog} Sh. M Kogan and A. Ya. Shul'man, Zh. Eksp. Teor. Fiz. 
{\bf 56}, 862 (1969) [Sov. Phys. JETP 29, 467 (1969)].
\bibitem{Altsh} B. L. Altshuler, L. S. Levitov, and A. Yu.
Yakovets, Pis'ma Zh. Eksp. Teor. Fiz.,
{\bf 59}, 821 (1994) [JETP Lett. {\bf 59}, 857 (1994)].
\bibitem{Korotkov} A. N. Korotkov, Phys. Rev. B {\bf 49}, 10381
(1994). 
\bibitem{Kor et al} A. N. Korotkov, D. V. Averin, K. K. Likharev,
and S. A. Vasenko, in {\it Single-Electron Tunneling and
Mesoscopic Devices,} edited by H. Koch and H. L\"{u}bbig
(Springer-Verlag, Berlin, 1992), p. 45.
\end{thebibliography}
\end{document}